\documentclass[prd,aps,twocolumn,preprintnumbers]{revtex4}
\usepackage[dvips]{graphicx,color}
\begin{document}\title{Charges on Strange Quark Nuggets in Space}


\author {E. S. Abers$^1$, A. K. Bhatia$^2$, D. A. Dicus$^3$, W. W. Repko$^4$, D. C. Rosenbaum$^5$, V. L. Teplitz$^{2,5}$}
\affiliation{$^1$Department of Physics and Astronomy, UCLA, Los Angeles, CA 90095-1547 \\$^2$NASA Goddard Space Flight Center, Greenbelt, MD 20771 \\$^3$Physics Department, University of Texas, Austin, TX 78712
 \\$^4$Department of Physics and Astronomy, Michigan State University, East Lansing, MI 48824\\\\$^5$Physics Department, Southern Methodist University, Dallas, TX 75275}
\date{November 8, 2008}

\begin{abstract}

Since Witten's seminal 1984 paper on the subject, searches for evidence of strange quark nuggets (SQNs) have proven unsuccessful.  In the absence of experimental evidence ruling out SQNs, the validity of theories introducing mechanisms that increase their stability should continue to be tested.  To stimulate electromagnetic SQN searches, particularly space searches, we estimate the net charge that would develop on an SQN in space exposed to various radiation baths (and showers) capable of liberating the SQN's less strongly bound electrons, taking into account recombination with ambient electrons.  We consider, in particular, the cosmic microwave background , radiation from the sun, and diffuse galactic and extragalactic ultraviolet backgrounds.  The largest charge,for the settings considered, develops on a solar system SQN exposed to a solar X-ray flare.  A possible dramatic signal of SQNs in explosive astrophysical events is noted.
\smallskip

\end{abstract}
\maketitle

\section{Introduction}

A generation has passed since Witten \cite{witten} suggested that strange quark matter (SQM) might be the ground state of all familiar matter and de Rujula and 
Glashow \cite{drg} set down a list of methods to search for strange quark nuggets (SQNs) in various mass ranges.  There have been 
significant efforts using most of those methods, but no SQNs have been found.  An area that has received, we believe,  little attention is the possibility
of exploiting the charge on SQNs in space that should be caused by radiation baths liberating electrons from the SQNs.  How challenging such exploration
would be depends on how large such charges might be.  We estimate those charges under various conditions, including those at the time of cosmic recombination, 
those in today's
cosmic microwave background  (CMB), those in the solar neighborhood from the quiet sun and from an X-ray flare, and those in the diffuse galactic and extragalactic radiation backgrounds.

We begin, in this section, with a brief review of SQM and SQN basics.  In Sec. II, we give the equations on which our numerical estimates are based and the 
approximations that go into deriving them.  We make simple, conservative approximations in estimating electron binding energies, wave functions and cross sections for large and small SQN masses and radii.  In a separate publication, we will show, in more detail than needed for the first estimates here, the wave functions in the transition between these two regions \citep{pip}.  In Sec. III we give the numerical results. In Sec. IV, we briefly discuss the results and their implications for the feasibility of space-based or space-directed electromagnetic SQN searches.  We conclude, in Sec. V, by collecting evidence, pro and con, on SQN existence to provide the reader a context in which to decide whether to devote time to exploiting our results for the charges on SQNs in space.

In 1984, Witten \citep{witten} (and  Bodmer \citep{bodmer} earlier) considered systems of up, down and ``strange'' quarks, pointing out that they would have the same attractive potential energy as systems of just up and down because the force between two quarks does not depend on their flavor.  They would, however, have about 10 percent less kinetic energy because the Pauli exclusion principle would not force the quarks into as high kinetic energy states as in the case with just two kinds of quarks.

Soon after this seminal work suggesting that SQM might be the lowest energy state of familiar, baryonic matter, Farhi and Jaffe \citep{fandj} 
worked out the basic nuclear physics of strange quark matter (SQM) within the MIT bag model \citep{mitbag}, and de Rujula \& Glashow \citep{drg} identified several 
observations that might lead to discovery of SQNs.  The latter proceeded from the basic relation for energy loss
{\bfseries
\begin{equation}
       dE/dt = -\pi r_N^2\rho_M v_N^3
\end{equation}}
where $v_N$ is the speed of the (spherical) nugget, $r_N$ its radius, and $\rho_M$ the density of the material through which it is passing.  
Equation (1) just says that the SQN must lose the energy needed to make the stuff in its way move as fast as it, the SQN, is moving.  
They estimated SQN mass ranges to which various sensors might be sensitive, for example the Earth network 
of seismometers being sensitive to masses over a 
ton or telescopes being sensitive to light emitted by SQNs entering the atmosphere.  They estimated the upper 
detectability boundary of mass regions by computing the mass M at which events would become too rare for the detector system if the galactic dark 
matter (DM) density, $\rho_{DM}$, were all in the form of SQNs of mass M.  Roughly, we have, for events per unit time,
\begin{equation}
\label{dndt}   dN/dt = v_N(\rho_{DM}/M)\pi r_d^2
\end{equation}
where $v_N$ is again the speed of the SQN, the dark matter density in the solar neighborhood is $\rho_{DM}=5 \times 10^{-25}$\, g\,cm$^{-3}$ and   $r_d$ is the radius of the detector system through which it is passing.  Equation (2) gives minimum detectable mass or speed in terms of the dark matter density limit on the abundance of SQNs of some one single mass. 

An important recent development is work by Alford \textit{et al.} \citep{wilcz} showing that, for SQM in bulk, 
Cooper pairing, the basic phenomenon of superconductivity should occur  with quarks.  The pairing could take
different forms.  For a useful review, see Alford \textit{et al.} \citep{alrev}  Most likely at high density, 
perhaps, would be pairs of quarks with equal and opposite momentum, antisymmetric in spin, flavor and color: color-
flavor locked (CFL) pairing.  The quark matter of such an SQN would be electrically neutral in its interior, but would have net positive charge on its surface, with the total (quark) surface charge proportional to the area. It would have quark charge $Z_Q=0.3A^{2/3}$ (where $A$ is one third the number of quarks) for an SQN with $m_s=150$\,MeV (see \citep{madsen}), balanced by electrons feeling a potential that is a step function at the SQN radial boundary and Coulomb beyond it.  We will discuss this model further below.

There are two potential sources for SQM. It might have been produced primordially, in particular in a phase transition in the early 
stages of the big bang. That was Witten's original thought, but it is, at best, controversial.  The issue is that an SQN formed at high temperature needs 
to cool. If it does that by evaporation, the SQN disappears.  Alternatively, it might cool by neutrino emission in which case it survives.  
There are experts and arguments on both sides (see \citep{bdvs} for a review and Section II.\,D below for a brief discussion).  Primordial SQNs would likely all be very close in mass assuming limited colliding after formation. The second potential source is ``neutron stars (NS).''  If SQM is the lowest energy state of matter, it is expected 
that Type II supernovae would likely rise to high enough temperature to cool into SQM.  If one did not, local quantum fluctuations would likely soon cause a 
global transition of the NS to an SQS (strange quark star).  Binary SQS 
systems would in time spin down, collide, and the galaxy would gain a population of SQNs of varying masses from the fragments.\\

\begin{table}
\noindent
\caption{Some Strange Quark Nugget Searches.}

\begin{ruledtabular}
\noindent \begin{tabular}{lll}

   Experiment/Observation&Mass Range (g) & Result \\
\hline
AMS\footnotemark[1] &$10^{-24} - 10^{-22}$ & not done\\
RHIC\footnotemark[1]&$< 3\times 10^{-21}$ & not found\\
Mica Tracks\footnotemark[2]&$10^{-20} - 10^{-14}$ &$<<\rho_{DM}$ \\
ICE CUBE\footnotemark[3]  &$10^{-3} - 10^{-2}$&not done\\
Seismometers:&&\\
\,\,\,Future Lunar \footnotemark[4]  &$10^{3} - 10^{6}$&not done \\

\,\,\,Apollo\footnotemark[5]&$10^4 - 10^6 $& $< \rho_{DM}/10$\\
  \,\,\,USGS Reports\footnotemark[3]& $10^6 - 10^8$&$< \rho_{DM}$ \\
\end{tabular}
\end{ruledtabular}
  \footnotetext[1]{Sandweiss\citep{ams}.}\footnotetext[2]{Price\citep{mica}.}\footnotetext[3]{Spiering\citep{ice}.}\footnotetext[4]{Banerdt \textit{et al.}\citep{betal}.}\footnotetext[5]{Herrin \textit{et al.}\citep{hrt}.}
\end{table}
There is significant, ongoing activity in searching for SQNs in various mass ranges.  See Finch \citep{finch} for a recent, brief review.  One important effort is the Alpha Mass 
Spectrometer experiment \citep{ams} which was due to be flown by the space shuttle to the space station in a couple of years. It would have been able to 
detect, 
and to distinguish from cosmic rays, light SQNs that would not penetrate the atmosphere.  However, the Columbia accident and the need to retire the shuttle
fleet and to complete the space station have prevented AMS launch as scheduled \cite{phystdy}.  A second is the ICE CUBE neutrino detector \citep{ice}
being installed in Antarctica.  It will have photomultiplier tubes to detect the products from collisions of (weakly interacting) neutrinos with the electrons
and nuclei of the ice.  It will also have acoustic detectors to be able to identify tracks made by SQNs.   It should be able to detect SQN masses up to as
much as about a gram. This is the highest mass for which Eq. (\ref{dndt})  gives a dozen or more events per year in a kilometer-sized region.
 
Past searches have included examination of tracks in mica by Price at Berkeley     \citep{mica}.  They have also included two cases of NASA work, in 2002, 
with evidence that two neutron stars were actually strange quark stars \cite{nasa}.  However, it was later concluded that alternate explanations for the observations were more 
likely \cite{oops}. 

Looking for evidence of SQNs was a prime objective of the Relativistic Heavy Ion Collider (RHIC) at Brookhaven National Laboratory. Since RHIC collides gold nuclei with gold nuclei, SQN masses up to two gold nuclei could, in principle, be produced. The experiment found none \citep{ams}.  However, the binding energy per quark of SQM would increase with increasing numbers of quarks \citep{fandj}, so it would not be surprising if systems with hundreds or even many thousands
  of quarks were not found, but larger assemblies were found.

Selected searches are summarized in Table I.  In the table, a few major searches are listed (we modestly include ours \citep{betal,hrt}), along with the mass range to which they will be/are/were sensitive and the result, where there is one, in terms of the inferred SQN density in our region of the galaxy. An interesting SQM space search using the equipment being deployed to monitor near Earth asteroids has been proposed by Horvath \citep{jhorv}.  Our methods below should apply to other, similar models such as that of Zhitnitsky \citep{zhit} on cold dark matter as compact composite objects. 

Our ultimate goal is electromagnetic space tests, in several different settings, designed to discover SQNs or to falsify current SQN models.  The near-term goal of this paper is to determine the charges that would develop on SQNs in these settings.   Each setting is characterized 
by a photon distribution and an ambient electron distribution.  In the next section we discuss the settings, models of SQN structure, and the formulary.  In the following
Section (III) we present our results.  In Section IV, we day dream of possible electromagnetic detection schemes.  In the final Section (V), we give a brief summary of evidence for and against SQM existence.

\section{The Formulary and Its Approximations}
\subsection{Settings} We consider extragalactic, galactic and solar system SQNs.  In each case, we specify photon and electron energy and angle (i.e.,radiation bath - isotropic, or unidirectional shower) distributions.   We go back in red shift z to recombination at z=1089..
\noindent
\begin{table}
\caption{\noindent Settings. The quantity $z_{III}$ is the red shift at formation of the first stars; $r_{SMBH}$ is the
radius of the supermassive black hole at the Milky Way center; (I)GDR is the (inter) galactic diffuse radiation and $r_{S}$ is the
solar radius. $T_0$ is the temperature of the quiet sun and $T_x$ of solar X-ray flux.}
 \noindent  \begin{tabular}{lllcr}\hline\hline    Location &&Radiation Source&&\\
&& \textit{Intergalactic} & 
\textit{Galactic} & \textit{Solar}\\
\hline
Intergalactic & & CMB: $(1+z)T_0$ & --- & ---\\
  &&+IGDR&&\\
&&&&\\
Galactic & &IGDR & GDR & ---\\
          & &$z_{III
}>z$  & $r>r_{SMBH}$&\\  &&&&\\
Solar & & IGDR &GDR & $r>r_S$\\
&&&&$T_0$\,or\,$T_x$\\
\hline \hline
\end{tabular}\\
\end{table}

In Table II we summarize these settings.  $r_{sc}$ is the
distance from the sun to the center of the galaxy,   $r_{SMBH}$ is the
radius of the supermassive black hole there,  and $r_{S}$ is the
solar radius.  For the cosmic microwave background (CMB)  we 
take just blackbody radiation at temperature $(1+z)T_0 = T$ where
$T_0=2.75$K, today's temperature.  Galactic diffuse radiation (GDR) and today's X-ray
intergalatic diffuse background radiation (IDBR) are both given by, \textit{inter alia},
 \citep{henry} and \citep{sands}.  Let the ionization of
the SQN be $Z_N$.  For each of the settings, we determine $Z_N$
by setting the rate of ionization $\dot{Z_+}$ by the photon bath
equal to the rate of electron capture $\dot{Z_-}$ from the ambient electron
bath. For each setting we need estimates for the speed and density
of free electrons (or even  hydrogen atoms in cases where $Z_N$ is
large enough to rip off  an atomic  electron) as well as for the energy distribution of the photon bath or shower.

SQN settings considered are the following. (1) For intergalactic  SQNs we have three settings: the CMB with $z=1089$ and $z=0$ as well as the IGDR with $z=0$ along with the dominant electron bath. (2) There are three settings for SQNs located in the solar system with charges determined by the competition between the photons of diffuse radiation, sunlight or X-ray flares and the electron shower of the solar wind.  (3) Finally, we also have SQNs at the  center of the galaxy (COG) subject to the local radiation and electron baths there.  

\subsection{SQN Structure Models} We consider cases with SQN structure such that the quarks are free to move (e.g. see \citep{wilcz}).  Then, 
quarks and electrons form Debye clouds which migrate to the surface, minimizing the energy of the system and leaving behind an electrically neutral system of quarks.  We make three major approximations in computing the binding energies of the electrons.  First, we approximate the potential energy of the least bound electron, in the case that $Z_N$ electrons have already been ionized, as in Fig. 1, that is

\begin{eqnarray}
\label{elpot}
	V_Z(r)&=& -Z_Ne^2/r,\,\,\,      r>r_N;\nonumber\\
	      &=& \infty, \,\,\,    r<r_N.
\end{eqnarray}

Second, where convenient, we replace the infinite barrier of Eq. (\ref{elpot}) and Fig. 1 by the potential of Fig. 2, an apparently violent approximation that doesn't actually make much difference at all except perhaps in the transition region, around nanogram masses, from atom-like SQNs (SQN radius less than Bohr radius) to large SQNs (opposite inequality) as discussed below.  The work of reference \citep{pip}  is designed to cover that transition region.   Third, we consider only s-wave electrons moving in the effective field of all their predecessors and the quark lattice.  This is a conservative approximation: by omitting angular momentum we are underestimating kinetic energy and hence underestimating the number of electrons that are ionized in the given setting.

These approximations are conservative with respect to our goal of identifying ways of detecting charged SQNs in space.  All three overestimate the amount of binding by omitting the centrifugal barrier, the electron-electron repulsion and the Pauli principle for the electrons. Therefore, they underestimate the amount of charge the SQN will acquire, and thus underestimate its detectability.

The result of the three approximations is to change the potential to that of Fig. 2.  The calculations below are based on that potential.

An important case is color-flavor locked (CFL) pairing \citep{wilcz}, in which the u, d, and s quarks pair symmetrically.  Then, everywhere in the bulk there are the same number of quarks of each kind and hence charge neutrality of the bulk of the quark lattice.  In this case the charge lies at the surface where the longer Compton wavelength of the (low mass) u and d quarks requires that the edge quarks be u and d unbalanced by any of the much more massive s quarks.  This gives a net total quark charge on the order of $Z_T\sim(M/m_p)^{2/3}$, where $m_p$ is the mass of the proton.  The full system has, in vacuo, in addition enough electrons so that the total charge of an isolated SQN is zero, and again the potential is that of Fig.\,1 which can be approximated here by that of Fig. 2.  The discussion in this paper is meant to apply directly to the CFL model.

 \begin{figure}
\includegraphics[width=7cm]{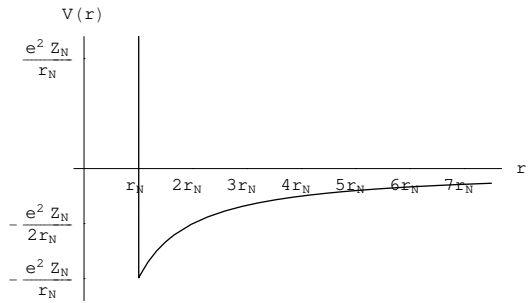}%
\caption{Potential for least bound electron.}
\end{figure}

\begin{figure}[tbp]

 \includegraphics[width=7cm]{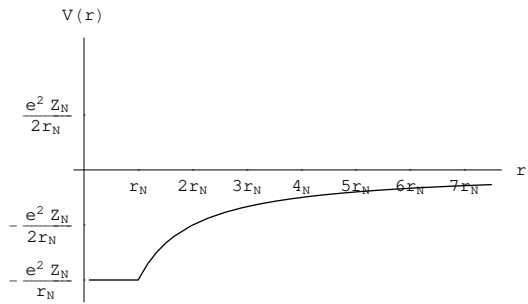}%
\caption{Approximation to potential for least bound electron.}
 \end{figure}

In Fig. 1, the cramped quarters might appear, by the uncertainty principle, to make the kinetic energy appreciable, therefore yield much less binding than the potential of Fig. 2, and hence give higher 
$Z_N$ values and easier detection.  However, the quarters are not really so cramped: the distance from $r=r_N$ to r such that $V(r)=V(r_N)/2$ is $r_N$.  This 
means that, for a given setting, the kinetic energy must fall with increasing SQN mass as $M^{-2/3}$ while the potential energy falls only as $M^{-1/3}$.  This occurs for the high M region, $r_N> a_B/Z_N$, with $a_B$ the Bohr radius for hydrogen.  In that case, we show below, Eqs.(\ref{fineq}) and/or (\ref{feq}) (or, numerically, in Figure 3), that $Z_N\sim b\sim M^{1/3}$.  If the kinetic energy becomes negligible, halving it doesn't much matter.

  The opposite case, low-mass atomic model, is $r_N< a_B/Z_N$.  There, the SQN is sufficiently small that the least bound electron is atomic in nature, in the sense that the overlap of its wave function with the SQN is small.  Thus, the effect of the third approximation is small.  We can get a sense of the effects of the other approximations by considering the variation in binding energy in the periodic table.  Our approximation treats all the SQNs (with $r_N< a_B/Z_N$) like hydrogen.  In real life, the most tightly bound last electron in the ground state of any atom in the periodic table is that of helium with binding energy about twice that of hydrogen while the least tightly bound, ground state electron is that of cesium with binding about one third that of hydrogen.   

There are multiple possibilities for SQN structure according to its surface tension (See Alford     \citep{surften}).  For surface tension below a critical value, large CFL SQNs decay into smaller ones while for surface tensions in the neighborhood of the critical value, an SQN develops a crystalline crust with smaller SQNs of sizes around a Debye length immersed in a gas of electrons.  We assume below that neither of these two cases obtain.  Rather, we assume: there are only CFL SQNs with surface tension sufficient to ensure a neutral baryonic bulk; and there is a uniform surface of u and d quarks retaining sufficient electrons to give total charge zero in the absence of external ionizing effects.  We then estimate, for different external settings, the charge that would develop on such an SQN. The upshot is that our results cover SQNs of masses for which SQNs may not exist or may not be stable. Even if SQM is the lowest energy state of (baryonic) matter and SQNs do indeed exist -- they are in other mass ranges. Recent work on SQN surface tension has been done by Oertel and Urban \citep{oerurb}

\subsection{The Formulary} We equate the 
rate, $\dot{Z_+}$ at which electrons are ionized by photons in the radiation bath (or unidirectional ``radiation shower'' in some cases ) to the rate at
which they are replaced by capture from the ambient electron bath or shower, $\dot{Z_-}$.  Our expression for $Z_+$ is

\begin{eqnarray}
\label{updot}
  \dot{Z_+}&=&\pi b^2 
\int^\infty_{Z_N e^2/r_N} dE\, N_\gamma(E)\nonumber \\
&& \left [\Big. N_e(E_B <E)\,
\sigma  (\gamma +SQN \rightarrow e+SQN),1\right]
\end{eqnarray}
where: $b= $max $[r_N,a_B/Z_N]$; $a_B$ is the conventional Bohr radius [$\hbar$c/($\alpha$mc$^2$)];
$r_N$ is (still) the SQN radius [$(3M/4\pi\rho)^{1/3}=r_N$]; $N_{\gamma}(E)dE$ is the
flux density of photons with energy between $E$ and 
$E+dE$;
$N_e$ is the number of electrons per unit area with binding energy $E$ or less;
$\sigma$ is the particle physics cross section for ionization 
and can be evaluated using standard relativistic quantum mechanics.  Here and below, $E_B=Z_Ne^2/b$ is the binding energy of the least tightly bound electron in the SQN.  The square bracket under the integral sign is 
the probability of the photon, when it hits the SQN, actually liberating an electron.  Because a probability is needed, when that product is greater than 
one, it must be replaced with one.

Similarly, we have

\begin{eqnarray}
\label{dndot}    \dot{Z_-}&= &\pi r_N^2\int^{\infty}_{m_e-E_B} v_e(E)n_e(E) \left[\Big. 
1+f_e(E,Z_N)\right]\nonumber \\ 
 &&h(E)g(e+SQN\rightarrow SQN+X, E)\, dE
\end{eqnarray}
where $h(E)$ is the distribution of incoming electron energies and $g$ is the probability that the incoming electron will be slowed and captured times the effective area (of the possibly macroscopic SQN) over which capture is possible.
We integrate over the distribution of incoming electron energies; $v_e(E)$ is 
electron speed; $n_e(E)dE$ is the number of electrons per unit volume with energies between $E$ and $E+dE$. The function $f_e$ is given by

\begin{equation}
\label{chfc} f_e(E)=\alpha\hbar cZ_N/(r_N E)\end{equation}
$f_e(E)$ is the enhancement of the
effective cross sectional area of the SQN due to the fact that the SQN  charge focuses the incoming electrons. It is the electrostatic analogue of the gravitational focusing factor discussed in detail on page 541 of Binney and Tremaine \citep{bandt}. It was introduced by V. S. Safronov \citep{saf} as an enhancement of the accretion rate of planetesimals in a dusty disk.   Note that the 
coefficient on the right-hand side in Eq.\,(\ref{updot}) is $\pi b^2$ while the one in Eq.\,(\ref{dndot}) is $\pi r_N^2$, (with $b$ the 
larger of $r_N$ and $\alpha \hbar c/(E_B Z_N)$). A photon can eject an 
electron out to $b$ while an electron must penetrate to $r_N$ to be captured in our approximation.  

Scattering off the bound  electrons 
or the much more numerous quarks and the absence of relativistic electrons at the Z values under consideration make it a reasonable approximation to assume that all  electrons up to total energies significantly higher than $m_e +E_B$ are 
captured.  In the spirit of this approximation we set the product, $hg$, equal to $\delta (E-\bar{E_e})$, where $\bar{E_e}$ is the average ambient electron energy, and 
replaced the whole integral by $n_e v_e(1+f_e)$.  This approximation would need to be revisited were we to consider an explosive event such as a supernova with MeV energies.

We solve the equation $\dot{Z_+}=\dot{Z_-}$ in the various settings described, letting M range 
from $10^{-21} \,\textrm{g} $ to $10^{+30}$ g.  We continue to make the conservative approximation that the kinetic energy is negligible compared to the 
potential energy for large $r_N$ and hence use, for the binding energy, $E_B\sim Z_Ne^2/r_N$ for $r_N> a_B/Z_N$.
 
For large $Z_N$, each SQN will be surrounded by an excess of unbound electrons and there will be screening that will affect both electron capture, Eq.\,(\ref{dndot}), by limiting the distance over which there is an attractive force, and electron liberation, Eq.\,(\ref{updot}), by impeding escape to infinity.  In the first approximation of this work, we do not attempt to estimate the size of these effects, but note that, because they have opposite effects, there is a possibility of first order cancellation.

We have solved the equation $\dot{Z_+}=\dot{Z_-}$ for some values of the parameters, and with some (further) approximations.  The most important of the approximations is replacing the product $N_e\sigma$ in Eq. (\ref {updot}) with unity.  This is a good approximation because, with $Z_T\sim M^{2/3}$, we have $N_e=Z_T/(4\pi b^2) \sim \,\textrm{few}\times10^{25}\mbox{cm}^{-2}$ --- independent of mass M. Since $\sigma\sim10^{-20}\mbox{cm}^2$, the probability of a photon liberating an  electron from an SQN is of the order of one if the photon's impact parameter lies within the effective cross section  from the center of the nugget, and if the photon is sufficiently energetic.  This approximation, $\pi b^2$ for the capture cross section, should be valid for electron energies (non-relativistic) of the settings in this paper (see Table III).

The final result of these approximations in Equations\,(\ref{updot}) and\,(\ref{dndot}) is that $\dot{Z_+}=\dot{Z_-}$ reduces to

\begin{equation}
\label{fineq}
		\pi b^2 cF_{\gamma}(E>E_B)=\pi r_N^2\bar{n}_e\bar{v}_e(1+\bar{f}_e)
\end{equation} 
where: $b$ is given just below Eq.\,(\ref{updot});
$F_{\gamma}(E>E_B)$ is the number of
 $\gamma$ 's per unit volume with energies greater than $E_B$; $E_B$ is the binding energy of the least bound 
remaining electron in the SQN; $n_e$ is the density of ambient electrons; $\bar{v}_e$ is their (average) speed;  
$\bar{f}_e$ is the (classical) focusing factor of Eq.\,(\ref{chfc});and barred quantities here and below are ensemble averages.   For four settings --- z=1089, z=0, quiet sun, and 
X-ray flare --- we use a thermal distribution.  For three --- intergalactic, near the sun, and milkyway center 
diffuse radiation backgrounds (DRB) --- we first used a non-thermal, power law spectrum approximating the X-ray spectrum graph \cite{henry}.  See also \citep{sands}. However, in the end  it was the UV spectrum which was important. The values are from page 143 of  Kolb and Turner \citep{kandt} . The number of photons per unit volume with energy $E_{\gamma}=E>E_B $, denoted F(E) is given by  

\begin{eqnarray}
\label{henryb}
	F(E) = -c^{-1}(\beta +1)^{-1}AE^{(\beta+1)}  
\end{eqnarray}
with $\beta=-3.05$; Log$(A)=7.1$ for intergalactic diffuse radiation spectrum.

Our approach to the SQN electrosphere is complementary to the classic treatment of Alcock, Farhi and Olinto \citep{afo}.  They consider the electrosphere as a continuous medium in the Thomas-Fermi approximation \citep{tfap}.  We consider the electron with the least binding energy in the mean field (of the positive lattice plus the remaining electrons) approximation.  Additionally, we do not make use of the fascinating work of Usov \citep{usov} and others \citep{usoth} on SQN cooling by thermal pair creation based on Schwinger's work on pair production in strong electromagnetic fields \citep{schwinger}.  We assume that the SQNs are cooled well below the temperatures at which thermal pair production becomes important.  Our consideration below of pair production is for a cold, even if highly charged, SQNs.

\section{Results} 
\begin{figure}
\includegraphics[width=3in]{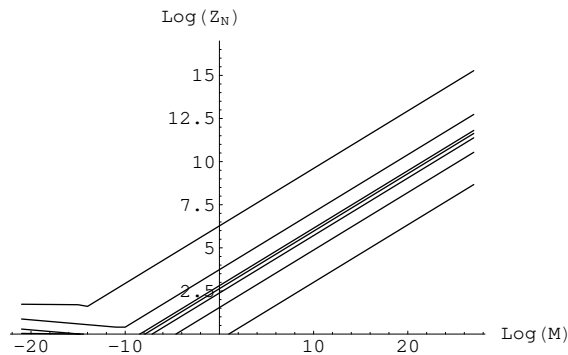}%
\caption{SQN charge $Z_N(M)$. From the top, the 7 curves are in the order described in the text at the beginning of Section III, and used for the 7 rows of Table III.   Mass, M, is in grams.}.
 \end{figure}

Figure 3 gives results for the equilibrium values of $Z_N$ for seven selected settings.  The seven curves, proceeding from top to bottom, are \\

 \noindent {\bfseries 1.}   the sun shining on an SQN near the Earth during an X-ray flare;\\
    {\bfseries 2.} the diffuse background radiation (DBR) shining on an SQN in the intergalactic medium (IGM); \\
 {\bfseries 3.}    the (quiet) sun shining on an SQN in solar orbit near the Earth;\\
 {\bfseries 4.}    an SQN in the primordial universe at recombination;\\
 {\bfseries 5.}     the galactic diffuse radiation (GDR) at the center of the galaxy (COG)
 shining on an SQN located near the center; \\
 {\bfseries 6.}      the GDR shining on a solar system SQN near the Earth; and\\
  {\bfseries 7.}     an SQN in the CMB today (ignoring the DRB and any other radiation).\\

In the Log-Log plot of Fig. 3, the third, fourth, and fifth	 curves from the top  are close to 
each other. The little, relatively flat tails on the left in Fig. 3.  give 
a rough approximation to the behavior in the mass region in which there is a transition between the ``atomic model'' 
($r_N < a_B/Z_N)$ and the $M^{1/3}$ behavior for $r_N > a_B/Z_N$. In Fig. 3 we have truncated curves where the computer program gives $Z_N(M)\leq 1$

Table III gives the parameters for the radiation and electron baths and showers in the same order as just described.  In two cases where parameters are unknown, we use our best estimates.  One is the electrons in the IGM.  We know that there about $10^{-6}$ baryons per cm$^3$ there.  Recent sightings of "warm-hot intergalactic matter" - WHIM (See, for example, Paerels \textit{et al.}  \citep{petal} for a brief review and further references) indicate potential abundance of WHIM in the IGM as high as fifty percent.  We assume ten percent in a rare burst of caution.  The energy distribution of the electrons in the WHIM is not known.  For the calculations, we take a generic 100\,K for it.  The second case where we estimate parameters is that of the center of the galaxy (COG), where conditions are not well known outside the immediate neighborhood of the supermassive black hole (SMBH).  We define the COG to mean the first 100\,pc.  This gives a volume large enough to be expected to have a significant number of SQNs, if they exist, as well as a number of baryons orders of magnitude larger than the number one would have from the SMBH's $10^6$ solar masses. but small enough to have approximately similar conditions throughout.  Assuming $r^{-2}$ behavior, there are $10^9$ solar masses of protons in the COG, (see  \citep{cox}) , and assuming 10 percent of that is ionized, gives us a value to use for the electron density. Again, we take a generic 100K for the electron temperature.

\begin{table}
\caption{Parameters for the seven curves of Fig. 3. E is in eV, $\bar{n}_e$ has units of cm$^{-3}$, $\bar{v}_e$ has units of cm/sec. F(E) is the function defined by Eq.\,(\ref{henryb}) for the non-thermal distribution cases. $F(E)=AE^{-3.05} \gamma/($cm$^2$\,s\,eV) where average temperatures are not known, we assume 100\,K.}
\noindent
\begin{tabular}{lllr}\hline
\hline
   SQN Location&Radiation &  $\bar{n}_e$&$\bar{v}_e/10^5$ \\

\hline
Solar X-ray  & $T=10^3$eV&$7 $&$500$\\
 \ \ Flare at 1\,AU \footnotemark [1]&&&\\
 IGM\footnotemark [2]&F(E); Log(A)=7.1 &$10^{-7}$&$60 $\\
 Near Quiet Sun \footnotemark[3] &$T=0.5$ eV&$7$&$500$\\
IGM\,Pre\,Recombo \footnotemark [4] &CMB $T=0.26$ eV&$287 $&$300 $\\
Galaxy Center \footnotemark [5]&F(E); Log(A)=11.57  &40&60\\
Near Sun(DRB) \footnotemark [6]& F(E); Log(A)=7.64&7&$500$\\
Today (CMB) \footnotemark[7]& $T=2.75$K&$10^{-7}$&$60$\\
\hline
\hline
\end{tabular}
\footnotetext[1]{Ref. [\onlinecite{cando}] p. 423;  p. 409; p. 409. Here and below, 1st entry is radiation; 2nd electron density; 3rd electron speed. }\footnotetext[2]{Ref. [\onlinecite{kandt}] p. 143; Ref. [\onlinecite{cox}] pp. 662-3, assume 0.1 ionized H; assume $T=100$\,K.}. \footnotetext[3]{Ref. [\onlinecite{cando}] p. 396; p. 409; p. 409.}
\footnotetext[4]{Ref. [\onlinecite{benetal}]; same for electrons.}\footnotetext[5]{Ref. [\onlinecite{cox}] p. 570; Ref. [\onlinecite{cando}] p. 571 extrapolated to 100 pc; assume 
$T=100$\,K .}\footnotetext[6]{Ref. [\onlinecite{kandt}] p. 143; Ref. [\onlinecite{cando}] p. 409;  p. 409.}\footnotetext[7]{Ref. [\onlinecite{benetal}]; Ref. [\onlinecite{cox}] pp. 662-3 ;  pp. 662-3.}
\end{table}

Based on Table III, one can understand the order of the curves in Fig. 3. The largest SQN charge, among the settings considered, is in the X-ray shower from a solar X-ray flare.  The neighborhoods of supernovae and other explosive astrophysical events, however, might well be even better, depending on the extent to which assets like the Swift satellite are able to get sufficient data, in seeing SQN effects.  The CMB by itself is, today, the worst, but it is perhaps interesting that, for sufficiently massive SQNs, it would contribute.  

In Fig. 4, we give the times (in years), $\tau_{eq}(M)$, necessary to reach 
$Z_N$ [$\tau_{eq}(M) = Z_N/\dot{Z}_{+} = Z_N/\dot{Z}_{-}$] for (just) the dominant contribution in each location.  
The results
for $Z_N$ do not vary with distance from the sun because both the radiation and electron showers fall off like $r^{-2}$.  However, the times to reach $Z_N$
do vary with r like $r^2$.  The times are about a day and a half for a nanogram, and fall as $M^{-1/3}$ as shown in Fig. 4.  See the discussion of Equation (9) below for a derivation of this result.


Roughly, there are three thick lines in Fig. 4.  The middle line, with the intermediate times is composed of three individual lines hard to distinguish in a log-log plot. They are the  settings 1, 3, and 5 discussed above and are the three solar system settings (near the Earth).  The line width is from the sizes of $f_e$ which are, for cases 1, 3, and 5  respectively, $4.9 \times 10^4$, 17, and 0.91. This gives at $M=1$\,g, $ \tau = 3.9 \times 10^{-6}, 3.7 \times 10^{-6},$ and $2.2 \times 10^{-6}$ years.  Similarly, the two intergalactic  settings (settings 2 and 7), with DBR and CMB radiation, have $f_e =9.9 \times 10^3$ and 0.85 giving times of 33 and 20 years.  Finally, COG today and universe just before recombination (settings 4 and 5) have $n_e/v_e=[40$\,cm$^{-3}$/60km/s; 287/300] giving $f_e=[44; 31]$ leading to $\tau=[8.1 \times 10^{-8}; 5.5 \times 10^{-8}]$  This last case is ``coincidental'' as far as one can tell.

\begin{figure}[tbp]
 \includegraphics[width=3in]{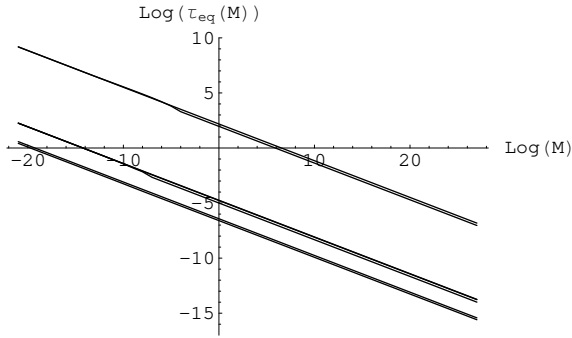}%
\caption {Time in years to reach equilibrium $Z_N(M)$. Top double line is IGM and CMB today; next, triple, line is the solar system at Earth's distance from sun during X-ray flare, from GDR, and for quiet sun. The bottom double line is  galactic center and universe at recombination.}
 \end{figure}

In the calculations we first took the DBR and the  galactic diffuse radiation (GDR) from Stecker and Salamon \citep{sands} combined with the graph of Henry \citep{henry}.  However, the result was binding energies only in the UV regions, so we switched to the reliable classic \citep{kandt} (page 143) to generate our results.  We said that the GDR around here (i.e. st the distance of the sun from the COG) could be extrapolated to the COG, which we took arbitrarily at 100 pc, by the $r^{-2}$ rule.

\begin{table}
\caption {Times and Binding Energies. The settings are grouped into three sections: the  three solar system ones together, the two intergalactic ones together and the universe just before recombination which is close for no apparent reason to COG.  These settings are not the same as in Fig. 3 and 4 or Table III.}
 \noindent
\begin{tabular}{lllr}\hline
\hline
   Setting & $M^{1/3}\tau_{Eq}$(y)&$E_B(eV)$&$E_B$(eV) \\
&&$M>10^{-10}$g&$10^{-21}$g\\
\hline
\hline
Galactic Center&$38 \times10^{-4} $&$4.5$&$52$\\
\,IGM Today: DBR&$151$&$100$&$760$\\
\hline
Solar system:&&&\\
\,\,during X-ray flare &$1.8\times10^{-5}$&$3.5\times10^4$&$3.9\times10^4$\\
\,\,from GDR &$1.0\times10^{-5}$&$ 0.65$&$12$\\
\,\,Quiet Sun &$1.7\times10^{-5}$&$12$&$18$\\
\hline
Recombo with CMB &$2.6\times10^{-7}$&$8.0$&$10$\\
\,Today from CMB &$92$&$8.7\times10^{-3}$&$0.012$\\
\hline
\hline
\end{tabular}
\end{table}
Along with Fig.\,4, in the second column of Table IV, we give results for the time $\tau_{Eq}(M)$ (in years) for $Z_N$ to reach equilibrium. Table IV goes on to give the binding energy of the least tightly bound electron, $E_B(M)$ for $M>10^{-10}$g and for$ M=10^{-21}$\,g --- for the seven settings.  Recall from subsection II.\,B above that $E_B=-Z_Ne^2/b;  b=$max$[r_N,a_B/Z_N]$; $10^{-10}$\,g is the boundary, roughly, between the two regions.    $\tau_{Eq}(M)$ varies as $M^{-1/3}$ for all settings.

Other variations with $r_N\sim  M^{+1/3}$ cancel out in the ratio $Z_N/\dot{Z}$:  Fig. 4 does not have the kinks that are present in Fig. 3.  The lack of any variation in $E_B(M)$ for $M>10^{-10}$ should aid in devising SQN detection schemes, in the case of fragments from collisions or other production mechanisms giving a distribution of masses.  The whole range would give the same signal.  In those cases in which the charge focusing parameter $\bar{f}_e$ of Eq.\,(\ref{chfc}) satisfies $\bar{f}_e>>1$, we can write a simple closed form for $\tau_{Eq}$.  Letting $b=$max$[r_N, a_B/Z_N]$, writing $\zeta=1(2)$ when $b \neq a_B/Z_N$ ($b=a_B/Z_N$), and recalling $E_B=Z_N \alpha\hbar c/(\zeta b)$, we have $\bar{f}_e=4\zeta bE_B/(r_NE_e)$.  Consequently, it follows from Eq.\,(\ref{chfc}) and the assumption that the 
integral in the capture rate, Eq.\,(\ref{dndot}), is $\bar{n}_e\bar{v}_e\bar{f}_e$ that we have
\begin{equation}
\label{eqtime}
\tau^{-1}=\dot{Z_N}/Z_N =\pi\alpha \bar{v}_e\bar{n}_e r_N(\hbar c/E_e)
\end{equation}
Eq.\,(\ref{eqtime}) shows that the time to reach equilibrium depends only on the electron bath (or shower) and the SQN mass, but not the radiation; this is confirmed in the numerical results of Table IV and Fig. 4.

We can go on to find a simple closed form equation for $E_B$, if $b=r_N$, in the approximations of either $\bar{f}_e \gg 1$ or $\bar{f}_e\ll 1$.  Using $\dot{Z}_{+}=\dot{Z}_{-}$ with Eqs.\,(\ref{chfc}) and (\ref{fineq}) gives, for the first of these two cases

\begin{equation}
\label{feq}
F(E_B)=\pi\zeta \bar{v}_e\bar{n}_er^2_N/(bE_e)
\end{equation}
where $F(E_B)$ is a function of a form that depends on the nature of the radiation distribution $N_{\gamma}$ (here, thermal or power law).  It is, for the case of $\bar{f}_e \gg 1$, the integral of the photon distribution function divided by $E_B$ from $E_B$ to infinity.

Variation of $E_B$ as a function of SQN mass $M$ is small for small $M$, as well as absent for large, in the case of thermal radiation.  This might be expected since the thermal spectral energy density cutoff is exponential (with $E_B$ in the exponent). But it is significantly less rapidly varying for the diffuse radiation backgrounds where it is only a low power and hence $Z_N$ is more sensitive to the electron bath or shower.. A similar separation of $\dot{Z}_+= \dot{Z}_-$ into an equation with one side depending only on the radiation and the other depending only on the electrons can be made if $\bar{f}_e\ll1$.  We do not provide estimates of $E_B(M)$ for $M$ in the transition region between small and large mass, since these should be most sensitive to more accurate wave functions of \citep{pip}.

Some features of the results include the following:

\begin{itemize}

\item  In Fig. 3 we have used the parameterization of Eq.\,(\ref{henryb}) for the diffuse background radiation, derived from that plotted in \citep{kandt} (intergalactic, in the area of the solar system same order of magnitude, and $r^{-2}$ extrapolation to the COG).    These are the only three non-thermal baths (or showers) considered.  
\item The fact that $Z_N(M)$ behaves as $M^{1/3}$, for the region in which $b=r_N$, i.e. where $M$ is large enough that the Bohr radius is inside the SQN, can be seen from Eq. 6.  If $b=r_N$ holds, all explicit dependence on M disappears from that equation and one just solves once for $E_B=\alpha \hbar Z_N/b$, the binding energy of the most loosely bound electron.  That the dependence of $Z_N$ on $M$ is the same as that of b implies $Z_N\sim M^{1/3}$.  
\item Physically, one expects that the value of $Z_N$ should rise with M so as to make the binding energy of the most loosely bound electron independent of the radius of the SQN in the M region for which $\pi b^2=\pi r_N^2$.  The more interesting question is that of the negative slope of the little tails on the left in Fig. 3.  (The answer is that, in that $M$ region, $\dot{Z}_-$ grows as $M^{2/3}$ from the increase in $r_N$, while $\dot{Z}_+$, with $b$ fixed at the appropriate Bohr radius, lacks that increase.  This drives the solution to Eq.\,(\ref{fineq}) to lower $Z_N$ to compensate.) 
\item The largest $Z_N$ values occur for solar system X-ray flares.  Since these only last for minutes, one will need to consider carefully whether there is sufficient time to realize the large values and, even more importantly, sufficient time to exploit them for SQN detection.  One sees, from Fig. 4, that the SQN mass must be over a ton before the time to reach equilibrium gets down to minutes.
\item The results for SQNs at recombination are about the same as the results for the (quiet) sun because, while the temperature in the sun is higher than  that in the CMB just  after recombination, the electron density then was lower than that in today's solar wind near the Earth.
\item Extragalactic SQNs have relatively high $Z_N$ values and ones near the solar system much lower ones because of the great difference in electron densities.
\item Note that $Z_N$ would be limited by rapid vacuum breakdown to less than $E_{B}(Z_N)=-2m_e$.  For $E_B<-2m_e$, it is energetically favorable for the vacuum to create an electron with $E_B=-2m_e$ with one $m_e$ going to the mass of the electron and the other $m_e$ used to make a positron which goes off to infinity.  However, Madsen \citep{madsennew} has recently studied, in some detail, pair creation implications for SQNs.  He points out that the rate of pair creation can be very slow, because the positron must tunnel through a Coulomb barrier. From his Eq. (10) one sees that the  time for pair creation rises like $M^{1/3}$ and passes one second at SQN mass of about one gram. Thus the $2m_e$ gammas will likely be at a higher energy for which the ionization rate equals the electron capture rate.   For $10^{-20}$g$ < M < 10^{-18}$g, pair production occurs at $Z_N>137$.  For  $M<10^{-20}$g, there are not enough electrons for it to occur at all.    We discuss detection of the $2m_e$ line further in the last subsection of Section IV. Three of us, in a separate publication, \citep{drt} have precisely computed the critical Z, as a function of SQN radius, by computing the Dirac Equation wave functions for $E=-2m_e$ for two different charge distributions.    
\item Our results do not take account of the energy cost of charge separation or of the stabilizing force of surface tension.  The net effect of these two factors can be to fractionate large SQNs if surface tension is small or to disfavor small SQNs if it is large.  The literature does not yet seem to contain reliable calculations of these effects, or the surface tension itself. In this paper we have considered the charge that would result  for the full range of mass values, in spite of the fact that some values may turn out to decay into smaller SQNs rapidly or even to be precluded entirely.


\item  The Thomas-Fermi approximation is valid where the electron wavelengths are smaller than the distance, $r_N$, over which the quark electric field varies appreciably.  Thus the transition with increasing $r_N$ to the Thomas-Fermi regime might take place at an $r_N$ such that the electron wave function becomes appreciable in or near the quark nugget --- around $r_N \sim 10^{-10}$cm.  This is just one order of magnitude smaller than the Bohr radius of the least bound electron.  References \citep{madsennew} and \citep{drt} compute the transition to the Thomas-Fermi regime more precisely, the latter by explicitly solving the Dirac equation. 

\end{itemize}

\section{Discussion}

There are many detection techniques to consider once the order of magnitude of the charges on SQNs, and the binding energies of their most loosely bound electrons, are known.  One could look for emission lines from electrons being captured by SQNs.  One could look for absorption lines and/or absorption edges in radiation coming toward us from behind an SQN population.  One could try to detect the static charge along with the very small charge-to-mass ratio, of nearby SQNs passing through the solar system.  Finally there is the possibility of indirect detection by means of some astrophysical effect of a population of charged SQNs. Below we give a few examples of such lines of inquiry.  The main problem in looking for a signal will be the low SQN abundance given the number density limit $\rho_{DM}/M>n_{SQN}(M)$.  If we got lucky, there could be a large SQN shower from a relatively nearby and recent collision of the neutron stars in a binary system, but it would be difficult to advise, with a straight face, anyone to spend real money on such a possibility.

\medskip
\noindent\textit{Particle Detection Techniques}.  One could consider space-based particle detection techniques, including ones that might be based on the Moon.  The major capability needed is wide area coverage.  Consider the event rate

\begin{equation}
\label{Nev}        dN_{ev}/dt = n_{SQN}v_{SQN}A
\end{equation}
where $n$ and $v$ are the number density and speed of the SQNs and $A$ is the effective detector area.  Again, suppose all SQNs are of mass $M$ and that $n_{SQN}\sim \rho_{DM}/M$.  If we want to detect one SQN in the time $\tau$, we need (recall $\rho_{DM} \sim 5\times10^{-25 }\,\textrm{g/cm}^3$)  $A\tau/M>10^{17}$ assuming that $v_{SQN}\sim250\,\textrm{km/s}$, the galactic virial velocity.  If a square kilometer could be instrumented, nanogram SQNs might be detected at rates up to 100/s, and one-gram SQNs once a year.  Something to consider is whether SQNs entering the atmosphere would produce a characteristic scintillation signal detectable by ground or space-based telescopes \citep{roy}.

\medskip
\noindent\textit{Absorption and Emission Lines and Edges.} A second approach is to
search settings in which there is identifiable absorption or
emission.  One example would be to look for settings in which there
is sufficient high energy radiation to bring SQNs to a high enough
degree of ionization that pair creation ensues if another electron
is ejected.  At zero temperature, we expect that to happen when $E_B$, the binding energy of the least bound electron, is $2m_e c^2$. This follows simply from energy conservation $(E_B+m(e^+)+m(e^-)=0$). See Madsen \citep{madsennew} for discussion of other properties of low temperature SQN pair production

At such a point, there should be an emission line at
$E=2m_e c^2$ from electron capture into those states with energies near $E_B=-2m_e$.  This would result from a strong enough photon distribution to preclude  lower $E_B$ while, at the same time, vacuum pair creation prevented a larger one.  The upshot would be that any electron capture would produce a $2m_e$ photon, a very distinctive signal.  However, as noted above, Madsen shows that the time required to produce a minimum energy pair rises exponentially with $M^{1/3}$ for SQNs, passing 1 second around a gram and shifting the $2m_e$ signal toward higher energy.  We emphasize that careful calculations are needed to determine whether that signal would be distinguishable against the strong backgrounds in an explosive event.   Once $E_B$ is at a high enough energy to permit rapid pair production, any further electron ejection
results in an $e^+\,e^-$ pair so that there would also be gammas with $E_{\gamma}=m_ec^2$ from $e^+-e^-$ annihilation, where the positron annihilation could be inside the SQN or outside the SQN with an ambient electron.  This signal should occur as the SQN, in the radiation and electron baths, oscillates about the equilibrium value of $Z_N$.  It requires an SQN mass above about $10^{-20}\,\textrm{g}$
so that the total number of electrons is sufficient to reach that
point. 

In addition to gammas from the positrons annihilating with ambient electrons and the $2mc^2$ emission
there might be an accompanying absorption edge at $2mc^2$ since photons of
energy over that value could liberate electrons from the SQN, but less energetic ones could not.  Our
preliminary calculations summarized above indicate the diffuse
radiation background is not able to ionize SQNs to that degree.
Other places to look would be at explosive events, including
supernovae (of various types), gamma ray bursts, and neutron star
bursts and superbursts.  

It is possible
that GLAST data, when available, could contain some evidence
of this triad of signals: an emission line at $E = m_ec^2$, an
emission line at $E = 2m_ec^2$, and an absorption edge starting from
$E = 2m_ec^2$.  The emission line at a single photon energy of $2m_ec^2$ (resulting from $\gamma$'s keeping $E_B\geq 2m_ec^2$ and vacuum pair creation keeping $E_B\leq 2m_e c^2$) would be a dramatic SQN signal.  It may well be that the background would swamp any
signal, but this three-pronged test could be such a definitive
indicator that it is important to assess its feasibility.  We know of no other source of $2m_ec^2$ emission lines.  A place to look for an analogous signal, without the emission line, would be in the solar system during a solar X-ray flare.  Then we would expect an X-ray signal of about 50 KeV and some absorption of X-rays over that (see Table IV and Fig. 4).  Additionally, we note that a strong X-ray signal from electron-positron annihilation strongly concentrated at the center of our galaxy has been observed most recently with the INTEGRAL satellite.  See, for example, Yuksel \citep{yuksel} for a brief review.  We have, however, not seen any reports of 1.02 MeV photons, so it would be premature for SQNs to clamor for entry into Yuksel's list of about two dozen ``exotic'' models that might account for the positrons.

A second place pair production is thought to have occurred was in pair instability supernovae \citep{fandh}.  At least one recent supernova, 2006GY, discovered by Quimby \textit{et al.} \citep{quimby}, and studied by Ofek \textit{et al.} and Smith \textit{et al.} \citep{ofek} showed strong positron production.  It was the most luminous supernova recorded and had a light curve that lasted significantly longer than the typical supernova.  Leahy and Ouyed \citep{leahy} have explained its features on the basis of a two step process, the first an ordinary type II supernova implosion followed by a transition from a neutron star to an SQS.  A second explanation might conceivably be possible using results along the lines of Fig. 4. A cloud of gram range SQNs from an event with persistent production of a lot of MeV range $\gamma s$ would generate positrons.  It could also generate the $2mc^2 \gamma$ signal discussed above and not expected from any other obvious, at least to us, processes.  This could be the proverbial ``smoking gun'' for the presence of SQNs. Pair production as a cooling mechanism for hot SQSs was pointed out by Usov  \citep{usov} and is an active area of study \citep{usoth}.

\medskip
\noindent\textit{Other Settings.} In addition to the 7 settings treated here, and the explosive case cited, there are many other places where electrons can be stripped from SQNs.  One example is SQNs captured into the Earth's geomagnetic field.  Paulucci, Horvath and Medina-Tanco \citep{horv} have considered this setting with trapped electrons both ionizing and "de-ionizing" the SQN.  Another interesting setting would be an SQN passing through the sun.  There, collisions with photons and protons would ionize: protons are much more effective than electrons because of the larger mass: collisions with an SQN with the galactic virial velocity gives a fraction of an eV for electrons compared to a KeV for protons.  The balance between ionization by photons and protons and electron capture by collisions with electrons should give a relatively high ionization, especially if the still greater energies possible from collisions with helium are taken into account.  The question would be one of the distance from which such a charge could be detected.  Passage of an SQN through a giant molecular cloud might be of interest. 

 Another important setting would be passage of an SQN through metal, rock or other common materials.  Estimating the charge produced by passage through rock, for example, would help determine whether detection in a deep underground laboratory would be feasible as well as the sort of signal that other detectors might give if a highly charged, massive body transited them.

In summary, we have computed, for seven different settings, the charge that would accumulate on an SQN passing through a photon bath (or shower) with  ambient electrons. In Fig. 3 we give the resulting equilibrium charges as a function of SQN mass. The largest charge among the settings considered for an SQN passing through the solar system during a solar X-ray flare.  In Fig. 4 we give an estimate of the times required to reach the equilibrium values. In Table IV we give the corresponding equilibrium binding energies of the least bound electron. We give these results for SQNs of masses ranging from $10^{-20}$ to $10^{+30}$g.

\section{Are They There?}

It is now three dozen years since Bodmer's paper \citep{bodmer} and two dozen since Witten's \citep{witten}, but there has been no experimental or observational sign of SQM -- in spite of a reasonably vigorous program of searching.  One has to ask the questions both of whether the game is worth the candle and of what is the score anyway?

First the score. There are no astronomical sightings of strange quark stars (SQS's).  As noted, two reports of strange quark pulsars \citep{nasa} have gone away \citep{oops}.  Additionally, there are two sets of observations that appear to preclude the neutron stars observed being SQSs.  First, there is the glitch problem.  Pulsars, from time to time, suddenly change rotation frequencies.  This change is most easily interpreted as the result of starquakes in the less dense, more familiar matter on top of the neutron core.  It is difficult to explain these glitches with SQSs since there is a limit to the mass of a familiar matter crust that can rest on a SQM core \citep{glit} (although one SQM model appears to do so \citep{sqnsinels}).  Then there are the explosions from time to time on soft gamma ray repeaters (SGRs).  These are believed to be neutron stars: some bursts are quite large and are, of course, called superbursts. The X-rays from superbursts show a structure of quasi-periodic oscillations (QPOs) during the course of the burst.  QPO frequencies identified during a couple of relatively recent superbursts correspond to those expected from a thicker crust \citep{watts} than neutron drip permits an SQM core to maintain \citep{afo}.

Moving on to terrestrial (and lunar) searches, observations and experiments were noted in Section I and summarized in Table I.  A more thorough review is given in \citep{finch}, references therein and other papers from that conference.

Last but not least is theory.  Recognizing that Cooper pairing should add tens of MeV per baryon to SQN binding energies \citep{wilcz} greatly enhanced the argument for SQM existence.  On the other hand, while observations have been casting doubt on the existence of SQSs (and fragments of SQS - SQS collisions), theory has been casting doubt on primordial SQN production.  The current situation is summarized by Boyanovsky \textit{et al.} \citep{bdvs} along the following lines: 

There is an estimate that a primordial SQN must have mass M greater than $10^{20}$\,g in order for neutrino cooling, which scales as $M$, to be greater than evaporation cooling which scales like $M^{2/3}$ \citep{bdys}.  Furthermore, if the (first order) phase change to SQM occurs, as expected, when the strong interactions become strong at about 200 MeV, one would expect a mass of this same order -- give or take a factor of $10^3$ or so.  This is about the mass of a large comet.  Assuming the solar system formed from a light year of gas, at a density of about $5\times 10^{-25}$g/cm$^3$, the galactic dark matter density in our neighborhood, there would, on the average, be one SQN every $10^{15}$cm (60 AU).  There could be on the order of a billion SQNs in the solar system, some in the Oort cloud, some in the Kuiper belt, and some in the interiors of the sun and the planets, but not many in nearby orbit about the sun since orbit perturbations by Jupiter and other planets tend to drive comet-sized objects out of the inner solar system.  There should also be SQNs passing through the solar system if they exist and if they constitute the galactic dark matter.

These would seem clear indications for where to concentrate future searches for primordial SQNs, but there appears to be a fly in the ointment:  again the neutrinos.  By diffusion, they wipe out inhomogeneities over distances larger than the neutrino mean free path in the time it takes for hadron bubble nucleation and quark-gluon plasma reheating. Returning neutrinos cancel neutrino emission.

Are they there? Is the game worth the candle?  Finch says in \citep{finch} ``If it turns out that SQM is stable, the implications would be potentially tremendous not only for the resultant direct and indirect understanding of the strong interactions but also for the practical applications ranging from new materials...''  In brief, the answer to ``are they there?'' seems to be that it's a long shot, but to some worth trying.

Our hope is that finding the charge on SQNs in space will help devise space tests that either detect SQNs or narrow the mass windows permitted.

\section*{Acknowledgments}
VLT very much appreciates a number of very helpful conversations with Demos Kazanas, one with Floyd Stecker on reference \citep{sands} and one with M. Alford effects of Cooper pairing.  Dr. Alford kindly provided a number of very helpful comments on the first version of this paper.  DCR and VLT are also grateful to Jonathan Gardner for calling SN-2006gy to our attention, as well as C. Kilbourne for talking WHIM with us.
DAD was supported in part by the U.S. Department of Energy under Grant No. DE-FG03-93ER40757.  WWR was supported in part by the National Science Foundation under Grant PHY-0555544.

\end{document}